\documentclass[conference]{IEEEtran}
\IEEEoverridecommandlockouts
\usepackage{cite}
\usepackage{amsmath,amssymb,amsfonts}
\usepackage{graphicx}
\usepackage{textcomp}
\usepackage{xcolor}
\usepackage[utf8]{inputenc}
\usepackage{algpseudocode}
\usepackage{hyperref}
\usepackage{algorithm}
\usepackage{algpseudocode}
\def\BibTeX{{\rm B\kern-.05em{\sc i\kern-.025em b}\kern-.08em
    T\kern-.1667em\lower.7ex\hbox{E}\kern-.125emX}}

\begin{document}

\title{Distribution Hub Optimization: Application of Conditional P-Median Using Road Network Distances}

\author{\IEEEauthorblockN{Faizan Faisal}
\IEEEauthorblockA{\textit{Department of Computer Science}\\
\textit{Lahore University of Management Sciences}\\
Lahore, Pakistan \\
Email: 23100030@lums.edu.pk}
\and
\IEEEauthorblockN{Zubair Khalid}
\IEEEauthorblockA{\textit{Department of Electrical Engineering}\\
\textit{Lahore University of Management Sciences}\\
Lahore, Pakistan \\
Email: zubair.khalid@lums.edu.pk}}

\maketitle

\begin{abstract}
This paper explores a GIS-based application of the conditional $p$-median problem (where $p=1$) in last-mile delivery logistics. The rapid growth of e-commerce in Pakistan has primarily benefited logistics companies, which face the challenge of resolving inefficiencies in the existing infrastructure and scaling effectively to meet increasing demand. Addressing these challenges would not only reduce operational costs but also lower carbon footprints. We present an algorithm that utilizes road-network-based distances to determine the optimal location for a new hub facility, a problem known in operations research as the conditional $p$-median problem. The algorithm optimizes the placement of a new facility, given $q$ existing facilities. The past delivery data for this research was provided by Muller and Phipps Logistics Pakistan. Our method involves constructing a distance matrix between candidate hub locations and past delivery points, followed by a grid search to identify the optimal hub location. To simulate the absence of past delivery data, we repeated the process using the population distribution of Lahore. Our results demonstrate a 16\% reduction in average delivery distance with the addition of a new hub.
\end{abstract}

\begin{IEEEkeywords}
Conditional $p$-median, Facility location problem, Urban logistics planning, Distribution hub expansion, Last-mile delivery, Optimization, Geographical information system
\end{IEEEkeywords}

\section{Introduction}

With the rise in smartphone accessibility and improved internet connectivity in Pakistan, e-commerce has experienced rapid growth over the past decade. In addition, the advent of COVID-19 led to a 76\% increase in the number of e-commerce merchants—from 1,707 in 2019–20 to 3,003 in 2020–21. The e-commerce market is further projected to grow from \$7.7 billion in 2022 to \$9.1 billion by 2025, representing an annual growth rate of 6\% \cite{salik_issue_2022}. Pakistan's logistics market is estimated at \$35 billion annually, and global freight transport is expected to grow 2.6-fold by 2050 compared to 2015 levels \cite{oecd_itf_2019}.

Rapid urbanization in developing economies like Pakistan has led to the emergence of large, dense cities such as Lahore. This urban growth has placed considerable strain on infrastructure and transport planning, as seen in the sharp rise in vehicle ownership \cite{kin2017sustainable}. The resulting increase in traffic congestion, loss of green spaces, and higher vehicle emissions underscore the need for sustainable urban logistics \cite{shah}. Effective management of these challenges is crucial to ensure the efficient distribution of goods, maintain service reliability, and uphold an appropriate level of service \cite{galkin2019last}.

The $p$-median problem aims to identify the optimal locations for $p$ new facilities to serve a set of demand points, minimizing the distance between the demand points and their nearest facility. When $q$ facilities already exist in the area and the goal is to add $p$ additional facilities, the problem is known as the conditional $p$-median problem \cite{DREZNER1995525}. After determining the positions of the $p$ new facilities, demand will be served by the nearest facility, whether existing or new. In this context, the facilities are distribution hubs responsible for dispatching couriers for last-mile delivery, while past deliveries represent the demand points.

In this paper, we propose a method to optimize the location of one additional facility ($p=1$), given that three facilities already exist. This is referred to as the conditional 1-median problem, or $(1,q)$-median problem.
\section{Literature Review}

Much of the literature on facility location modeling has focused on formulating new models and modifications to existing models, striving to produce more efficient techniques \cite{current2002discrete}. Minieka \cite{minieka1980conditional} was the first to formally introduce the concept of the conditional location problem, focusing on studying conditional centers and medians on a graph. In 1990, Chen \cite{chen2020application} proposed a technique for addressing minisum and minimax conditional location-allocation problems when $p \geq 1$. Berman \cite{berman1990conditional} proposes a solution that solves the conditional 1-median problem by applying a $(p+1)$-median problem on an origin-destination matrix. This modified matrix includes a new hub $a_{o}$ and an arbitrary demand point $\upsilon_{o}$.

To solve the problem, Drezner \cite{DREZNER1995525} proposes a heuristic solution to the $(p,q)$-median problem ($p$ are the new facilities and $q$ are existing ones), which produces 29 local minima in 100 randomly generated starting solutions. The solution makes use of repeatedly solving the $p$-median problem and requires an algorithm that works on the original distance matrix.

In 2008, Berman and Drezner \cite{berman2008new} proposed a method to solve both the conditional $p$-median and $p$-center problems, which involves solving an unconditional $p$-median and $p$-center problem using the shortest distance matrix just once. Kaveh and Esfahani \cite{kaveh2012hybrid} introduced a hybrid approach, combining a harmony search and a greedy heuristic algorithm for addressing $p$-median problems.

Irawan et al. \cite{irawan2014adaptive} adopt a multiphase approach that incorporates demand point aggregation, Variable Neighborhood Search (VNS), and a proposed method for solving large-scale unconditional and conditional $p$-median problems. Chen et al. \cite{chen2020application} present a method for applying the $p$-median model to logistics node location, where the distance between demand points and the logistics center is calculated based on linear distance. In our methodology, we use distances derived from the original road network.

\section{Methodology}
\subsection{Mathematical Modeling}

Let the set of existing hubs be denoted by

\begin{equation}
Q = \{q_1, q_2, \dots, q_k\}, \quad \text{where } k = |Q|,
\end{equation}

and the set of delivery points be denoted by

\begin{equation}
N = \{n_1, n_2, \dots, n_a\}, \quad \text{where } a = |N|.
\end{equation}

Let the next optimal hub to be placed be denoted by \( p \). Furthermore, let \( d(x, y) \) represent the shortest distance between points \( x \) and \( y \).

Our objective is to find

\begin{equation}
\underset{p}{\arg\min}\; F(p),
\end{equation}

where the objective function \( F(p) \) is defined as

\begin{equation}
F(p) = \sum_{i=1}^{a} \min \left( \underset{1 \leq j \leq k}{\min}\, d(q_j, n_i),\; d(p, n_i) \right).
\end{equation}

\subsection{Implementation}

\begin{figure}[htbp]
\centerline{\includegraphics[width=\linewidth]{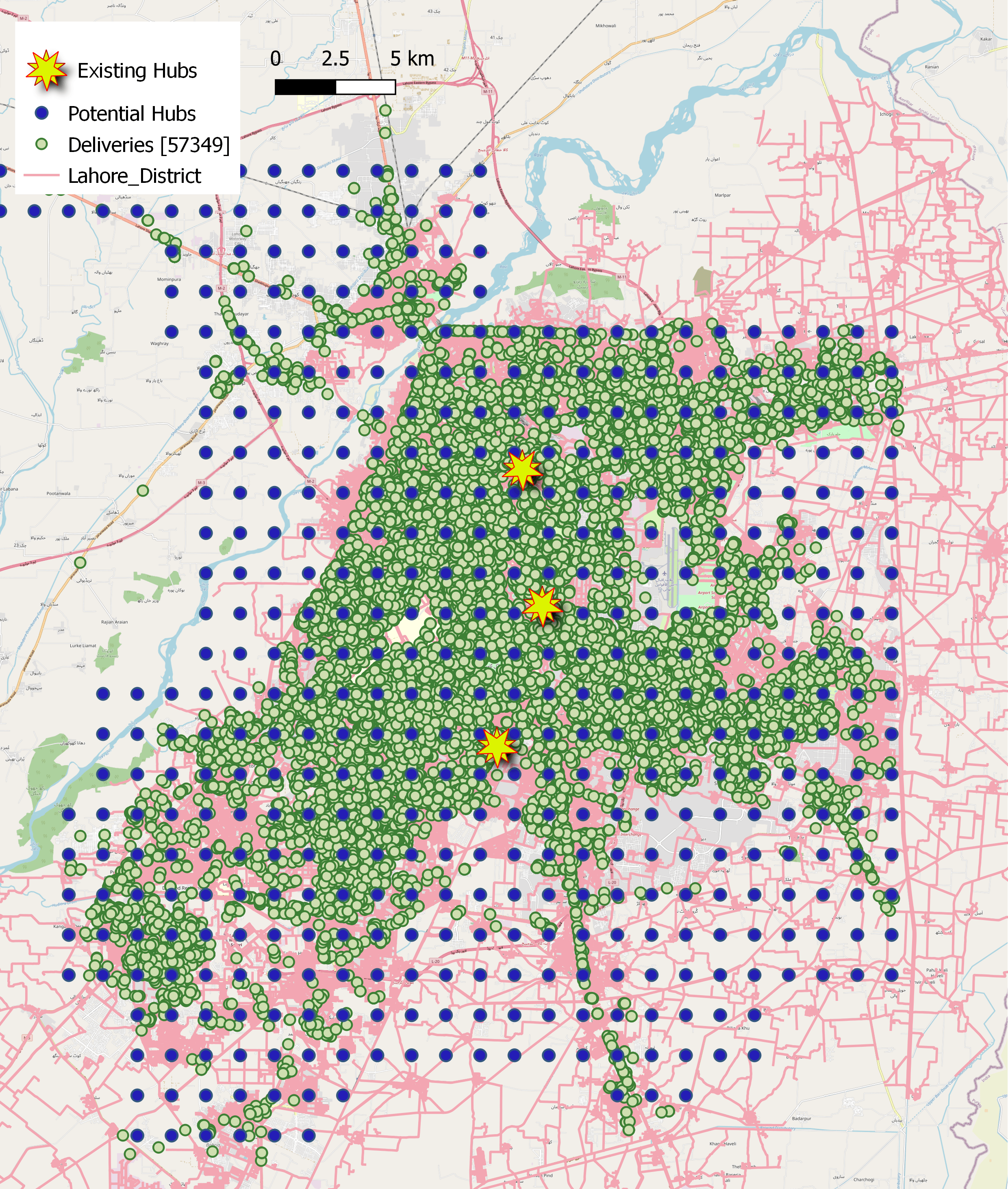}}
\caption{Potential locations for the 4th hub in Lahore}
\label{fig:potential_hubs}
\end{figure}

We placed candidate hubs approximately 1.3 km apart from each other. There were three existing hubs and 491 candidate hubs, making a total of 494 hubs. These 494 hubs served as the source points. Figure \ref{fig:potential_hubs} shows all potential locations for the 4th hub on the map of Lahore.

\begin{figure}[htbp]
\centerline{\includegraphics[width=\linewidth]{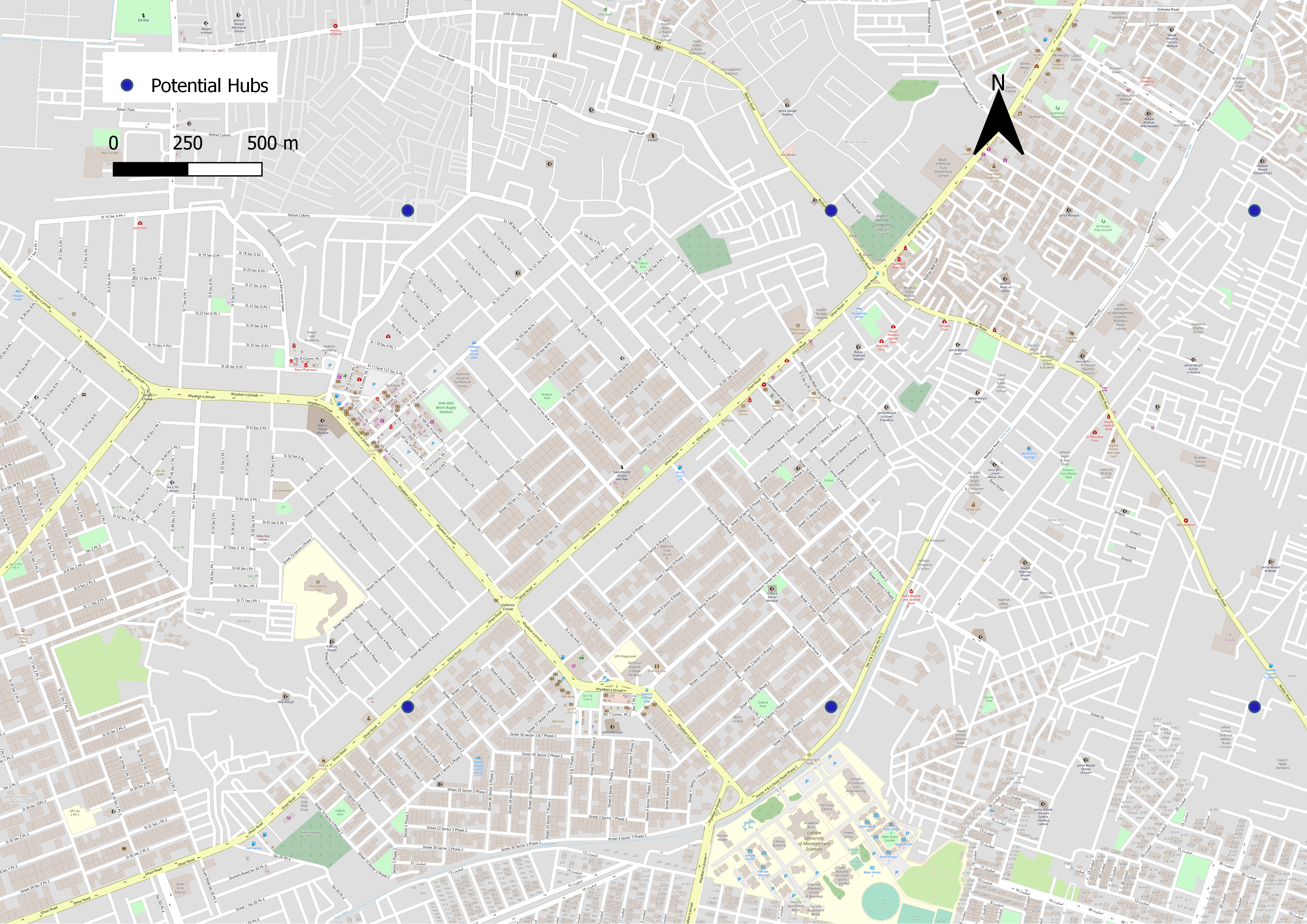}}
\caption{Zoomed-in view of potential 4th hub locations}
\label{fig:hub_scale}
\end{figure}

Figure \ref{fig:hub_scale} provides a close-up view of the potential locations for the 4th hub, illustrating the distance between each successive point.

\begin{figure}[htbp]
\centerline{\includegraphics[width=\linewidth]{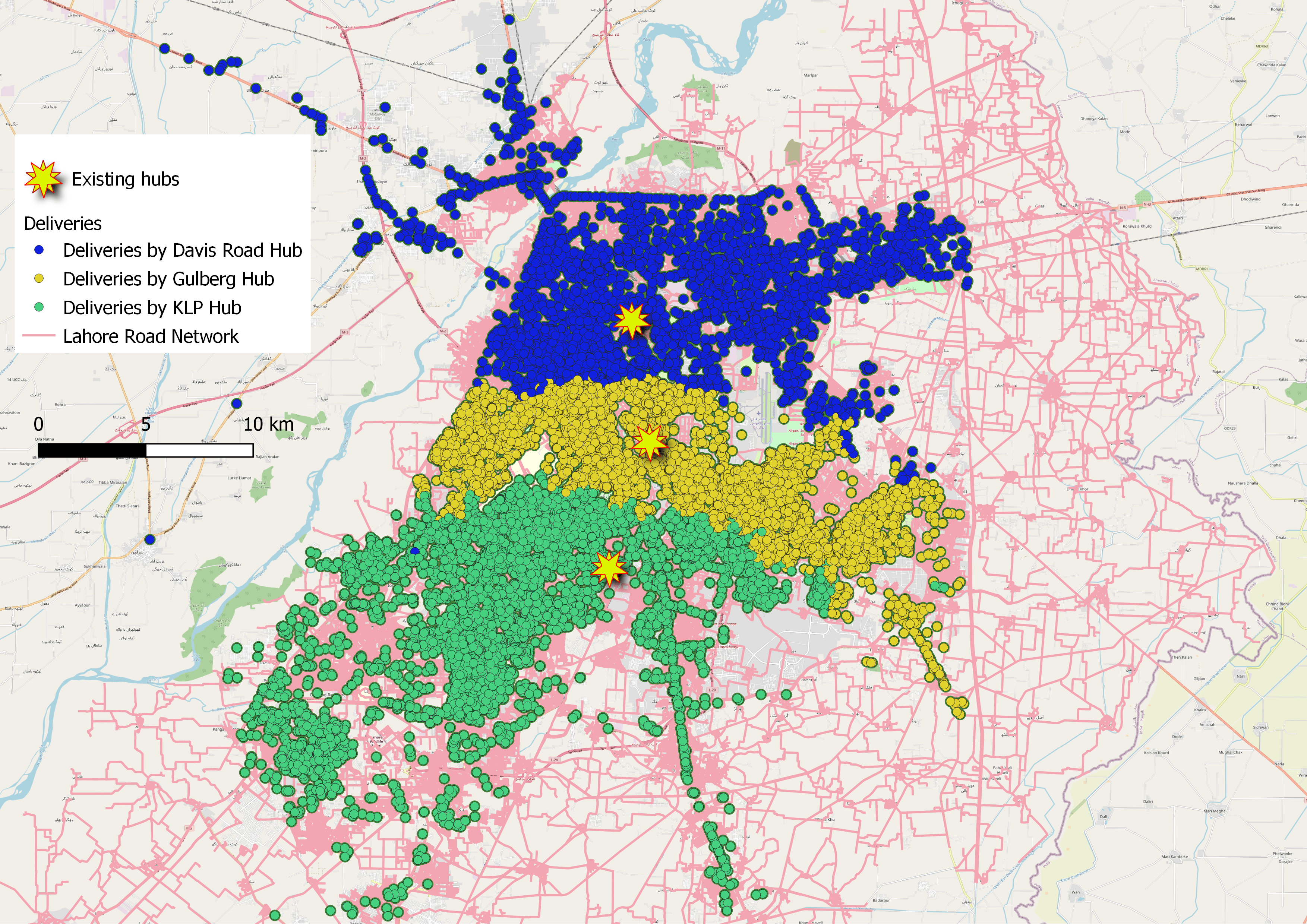}}
\caption{Existing 3 hubs and geospatial past deliveries data in Lahore}
\label{fig:existing_hubs}
\end{figure}

A random sample of 20,000 deliveries was selected from the cleaned dataset of 57,349 total deliveries to analyze the distribution of deliveries. The sampling process was performed using the Pandas \cite{noauthor_pandas_nodate} library in Python \cite{noauthor_welcome_2024}. Figure \ref{fig:existing_hubs} illustrates the locations of the three existing hubs along with the geospatial data of past deliveries on the map of Lahore.

To evaluate the optimal candidate hub, a distance matrix $D_{i,j}$ was constructed where $i$ represents the delivery point (destination), and $j$ represents the hub (source). This distance matrix was constructed using the Openrouteservice API \cite{noauthor_openrouteservice_nodate} in QGIS \cite{noauthor_spatial_nodate} and Python.

The API leverages OpenStreetMap \cite{noauthor_openstreetmap_nodate} and the Contraction Hierarchies (CH) algorithm \cite{geisberger2008contraction} to compute distances between source and destination points via road networks. The CH algorithm is a speed-up technique that enhances Dijkstra's algorithm \cite{dijkstra2022note} by skipping non-essential vertices when calculating the shortest path between two points in a graph (in this case, the road network).

\begin{figure}[htbp]
\centerline{\includegraphics[width=\linewidth]{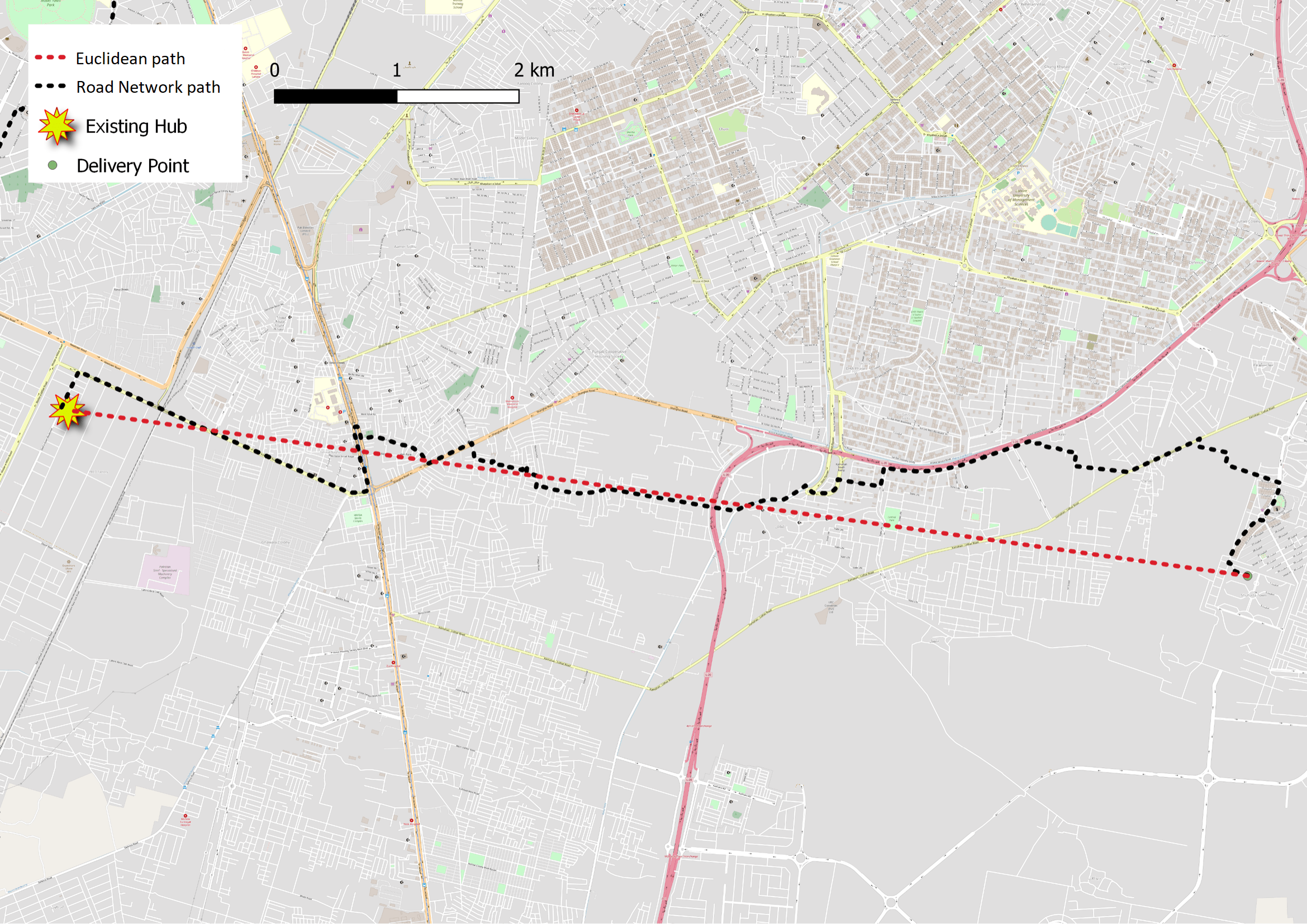}}
\caption{Comparison of road network distances vs Euclidean distances.}
\label{fig:road_vs_euclidean}
\end{figure}

Figure \ref{fig:road_vs_euclidean} presents a comparison between road network distances and Euclidean distances, emphasizing the significance of using road network distances for achieving greater accuracy.

Using the API, a distance matrix of dimensions 20,000 (deliveries) × 494 (hubs) was constructed.

Subsequently, our algorithm was applied to the distance matrix, with the indices of the existing hubs as an additional input. The algorithm returned the lowest cost and the matrix index (column number) of the globally optimal hub. This index was then used to retrieve the corresponding coordinates of the optimal hub.

The deliveries were re-clustered for the $q+1$ hubs, and the results were visualized. This procedure was repeated for $q=2$ (Davis Road and Gulberg hubs), which effectively relocated the KLP hub to an optimal position, conditioned on the locations of the other two existing facilities.

By iterating over all candidate hubs $(494 - q)$, where $q$ is the number of existing hubs, our algorithm minimized the total distance by assigning each delivery to its nearest hub.

The cost function minimized was the total distance covered (in meters) in delivering 20,000 orders by $q+1$ hubs. A summary of the algorithm is provided in Algorithm 1. The time complexity of the algorithm is $O(nm)$, where $n$ represents the number of hubs (both candidate and existing), and $m$ is the number of deliveries.

\begin{figure}[htbp]
\centerline{\includegraphics[width=\linewidth]{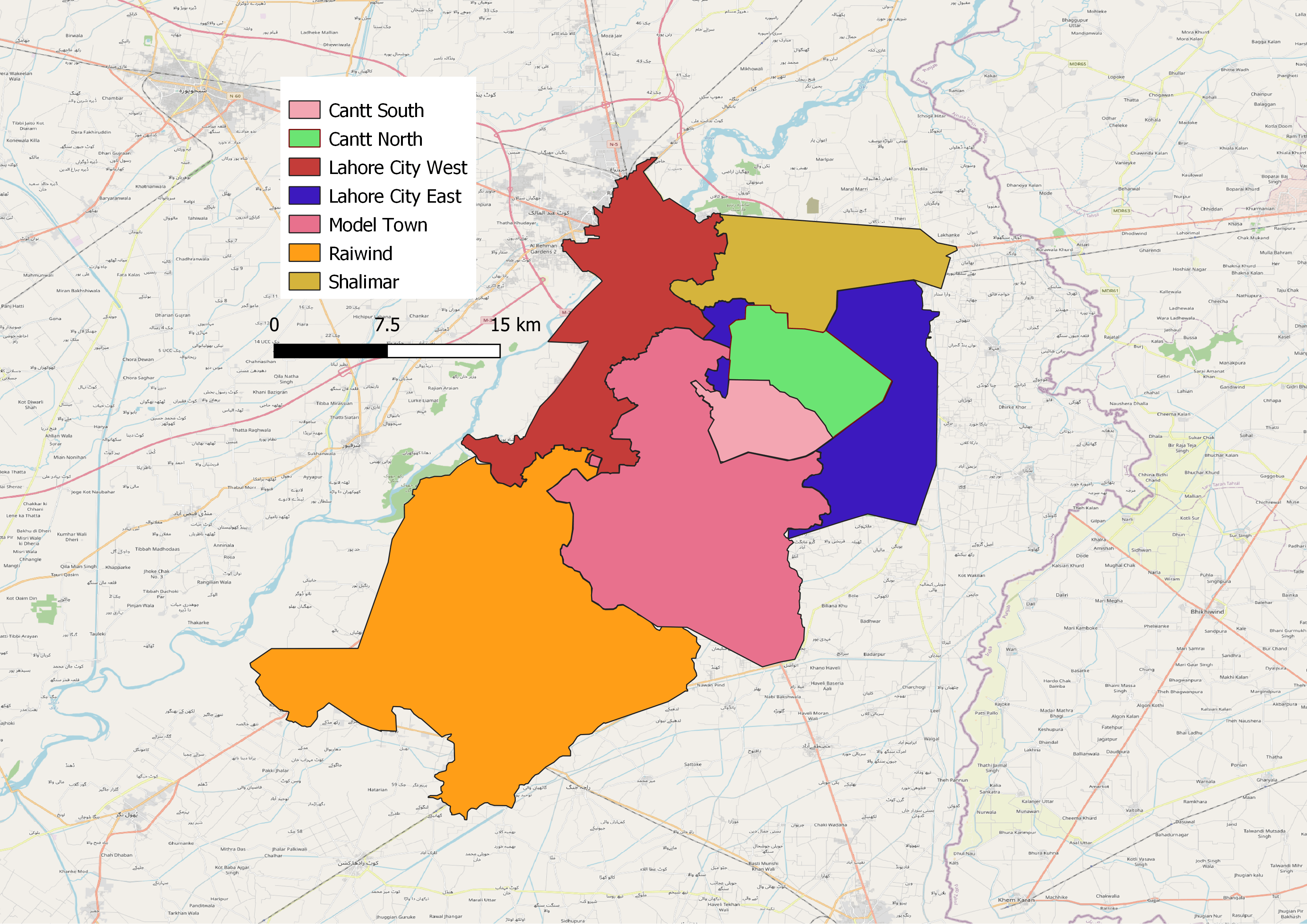}}
\caption{Different Tehsils (areas) of Lahore}
\label{fig:tehsils}
\end{figure}

To assess the significance of the delivery data, we repeated the process using the population data of Lahore. The city is divided into five tehsils: Cantt, Model Town, Lahore City, Shalimar, and Raiwind. Figure \ref{fig:tehsils} depicts the division of Lahore into its respective tehsils.

\begin{figure}[htbp]
\centerline{\includegraphics[width=\linewidth]{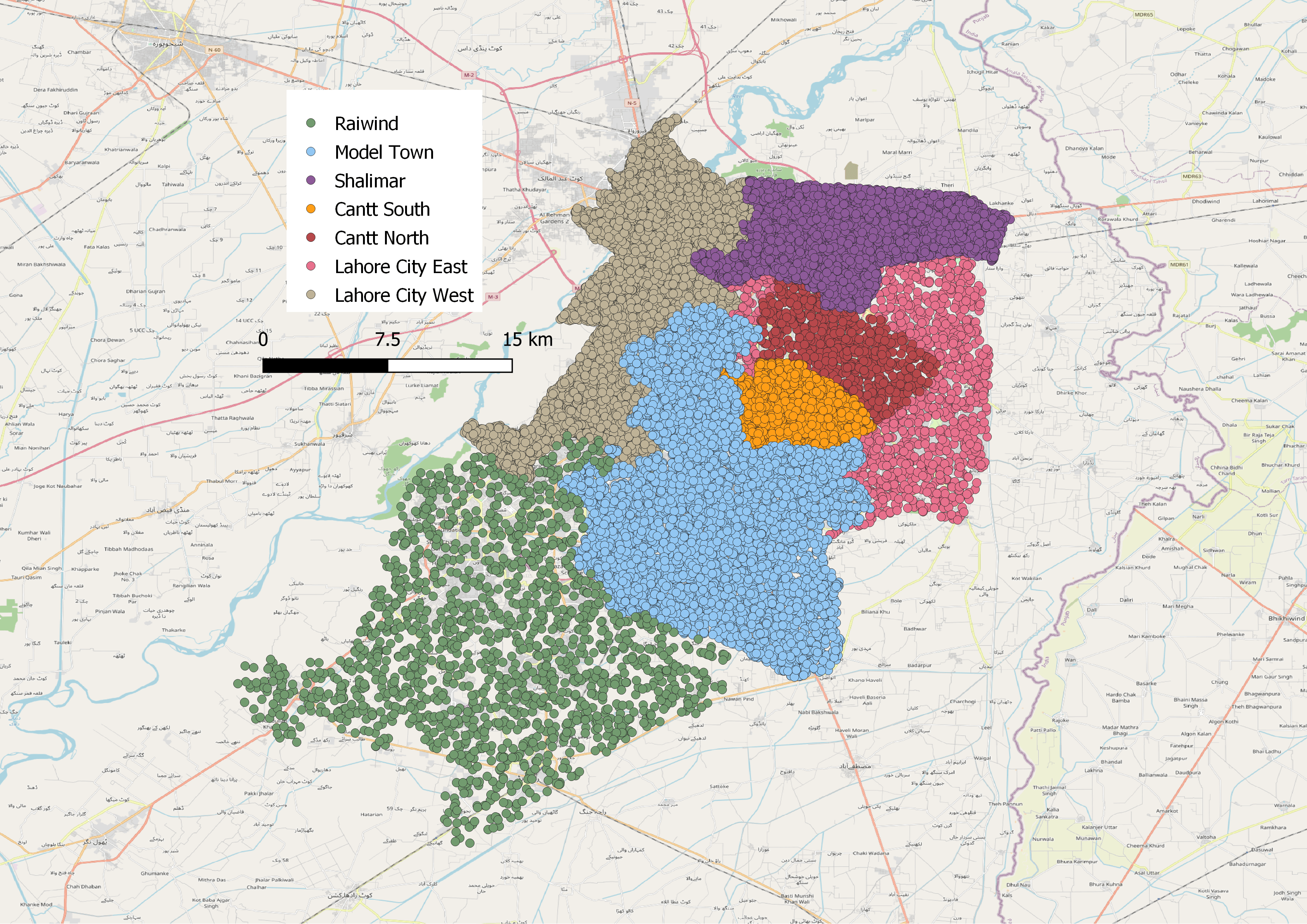}}
\caption{Population distribution of Lahore in different Tehsils}
\label{fig:population_tehsil}
\end{figure}

We used data from the Pakistan Bureau of Statistics \cite{noauthor_home_nodate} and the Election Commission of Pakistan \cite{noauthor_election_nodate} to determine the population strength of each tehsil. The total population was scaled to 20,000 people, and the population points were randomly distributed uniformly in each of the tehsils using the QGIS vector creation plugin (see Figure \ref{fig:population_tehsil} and Table \ref{tab:population_distribution}). Each population point essentially represented a demand point.

\begin{table*}[ht]
\caption{Summary of Results}
\label{tab:summary_of_results}
\centering
\begin{tabular}{|l|l|c|c|}
\hline
\textbf{Method Used} & \textbf{Data Used} & \textbf{Avg. Delivery Distance (km)} & \textbf{Improvement} \\
\hline
New hub              & Past deliveries    & 6.385                               & 16.0\%               \\
Relocating hub       & Past deliveries    & 7.229                               & 5.0\%                \\
New hub              & Population         & 6.785                               & 10.8\%               \\
\hline
\end{tabular}
\end{table*}

\begin{table}[ht]
\caption{Distribution of Population}
\label{tab:population_distribution}
\centering
\begin{tabular}{|l|c|}
\hline
\textbf{Tehsil}               & \textbf{Scaled Population} \\
\hline
Cantt (North)                 & 672                        \\
Cantt (South)                 & 1,273                      \\
Model Town                    & 4,863                      \\
Lahore City (West)            & 6,571                      \\
Lahore City (East)            & 992                        \\
Shalimar                      & 4,103                      \\
Raiwind                       & 1,526                      \\
\hline
\textbf{Total}                & \textbf{20,000}            \\
\hline
\end{tabular}
\end{table}

\begin{figure*}[t]
    \centering
    \begin{minipage}{\textwidth}
        \begin{algorithm}[H]
        \caption{Optimal Hub Selection}\label{alg:optimal_hub}
        \begin{algorithmic}[1]
        \State $DistanceMatrix \gets \text{Input}$ \Comment{Matrix of distances between hubs and delivery points}
        \State $ExistingHubs \gets \text{Input}$ \Comment{List of indices of existing hubs}
        \State $MinimumCost \gets \infty$ \Comment{Initialize the minimum cost to infinity}
        \State $BestHubIndex \gets \text{Null}$ \Comment{Index of the optimal hub to be determined}

        \For{$CandidateHubIndex$ in $DistanceMatrix$} \Comment{Iterate over potential hub locations}
            \If {$CandidateHubIndex$ in $ExistingHubs$} \Comment{Skip existing hubs}
                \State \textbf{continue}
            \EndIf

            \State $RelevantDistances \gets DistanceMatrix_{CandidateHubIndex, m}$ \Comment{Extract distances between candidate hub and delivery points}
            \State $CurrentCost \gets 0$ \Comment{Initialize cost for the current candidate hub}
            
            \For{$i = 0$ to $N-1$} \Comment{Assuming N is the number of existing hubs}
                \State $CurrentCost \mathrel{\gets} CurrentCost + \min\left(RelevantDistances_{i}\right)$ \Comment{Add the minimum distance for this hub}
            \EndFor
            
            \If{$CurrentCost < MinimumCost$} \Comment{Check if current cost is less than the minimum cost found so far}
                \State $MinimumCost \gets CurrentCost$ \Comment{Update the minimum cost}
                \State $BestHubIndex \gets CandidateHubIndex$ \Comment{Update the best hub index}
            \EndIf
        \EndFor
        \end{algorithmic}
        \end{algorithm}
    \end{minipage}
\end{figure*}

\section{Results}

\begin{figure}[htbp]
\centerline{\includegraphics[width=\linewidth]{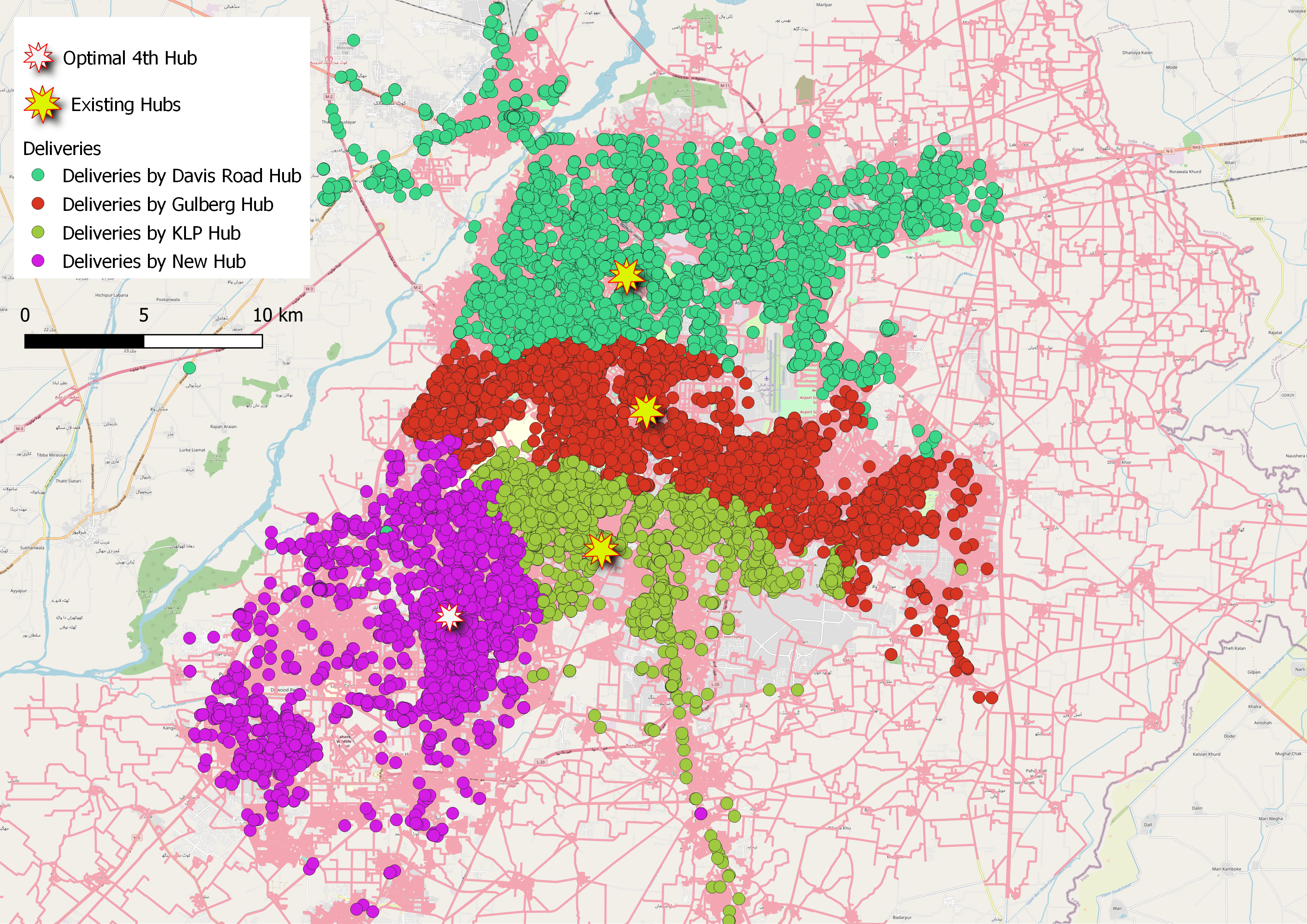}}
\caption{Optimal 4th hub location determined along with the existing 3 hubs on the map (deliveries data)}
\label{fig:new_hub_location}
\end{figure}

Figure \ref{fig:new_hub_location} illustrates the location of the newly established hub on the map of Lahore, determined using past delivery data. The average distance per delivery was selected as the metric to assess the improvement achieved through the addition of this hub. Consequently, the average distance per delivery decreased from 7.609 km to 6.385 km, representing a reduction of 16.0\%.

We observed that the most common delivery distances shifted from the range of 5–7 km to 3–5 km. Furthermore, the number of outliers in the distribution was reduced. The runtime of the algorithm on the distance matrix was recorded at 0.187 seconds.

\begin{figure}[htbp]
\centerline{\includegraphics[width=\linewidth]{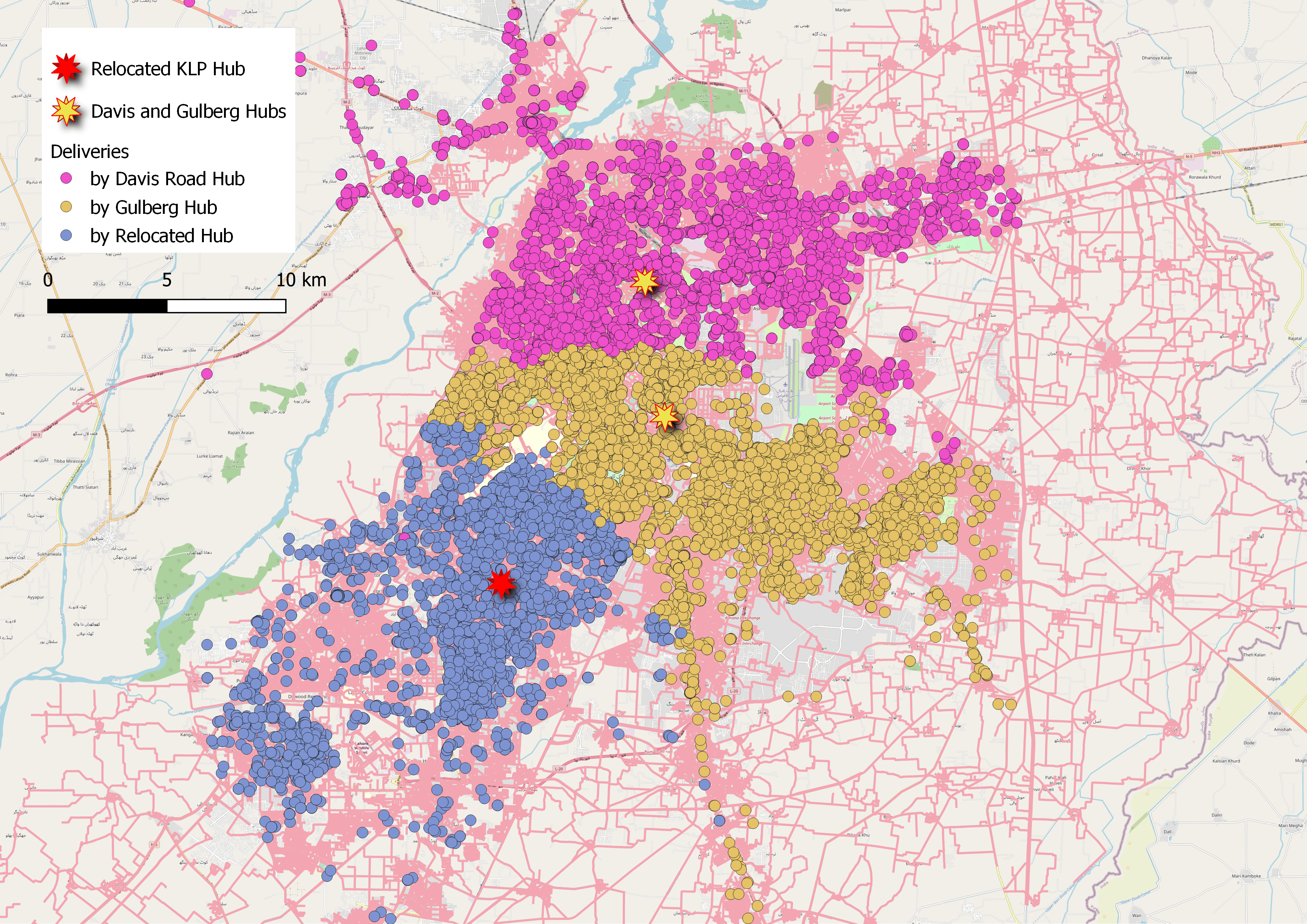}}
\caption{Relocated KLP hub on the map of Lahore}
\label{fig:relocated_hub}
\end{figure}

\begin{figure}[htbp]
\centerline{\includegraphics[width=\linewidth]{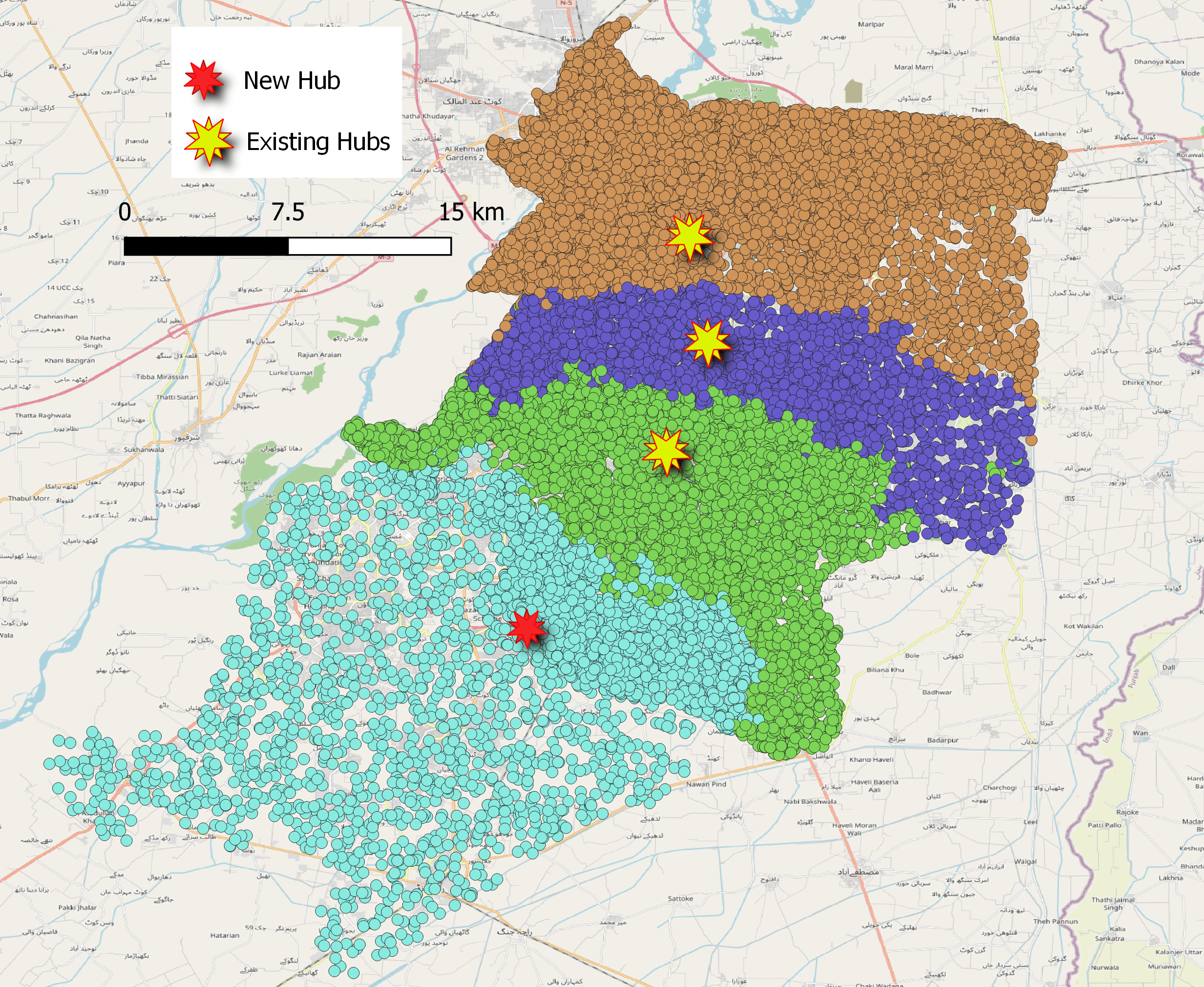}}
\caption{Location of the existing and new hubs based on population data}
\label{fig:population_hubs}
\end{figure}

\begin{figure}[htbp]
\centerline{\includegraphics[width=\linewidth]{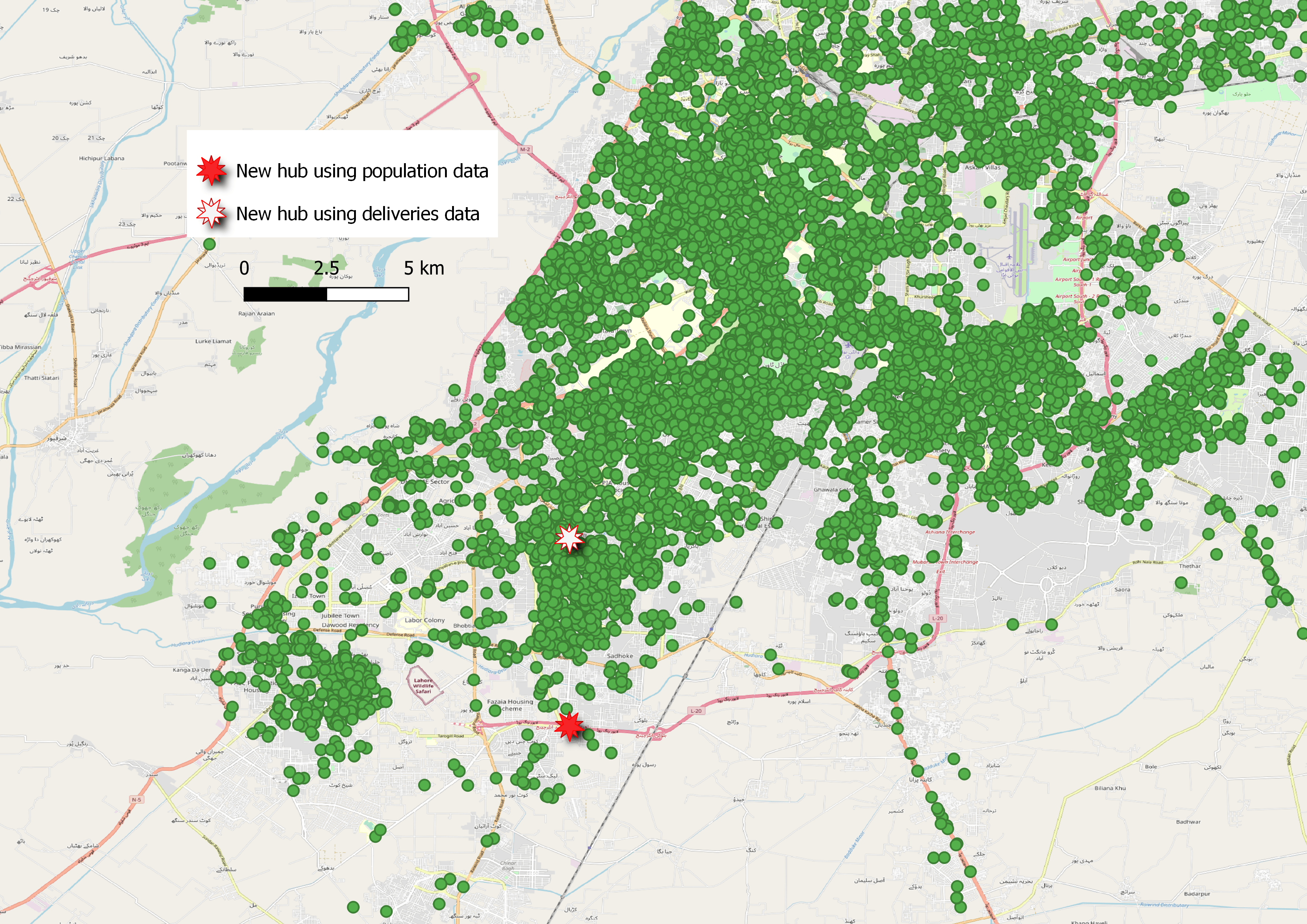}}
\caption{Comparison of hub locations determined via past deliveries data and population data}
\label{fig:hub_comparison}
\end{figure}

By conditioning on the Davis Road and Gulberg hubs, the relocated KLP hub resulted in a reduction of 5.0\%, bringing the average distance down to 7.229 km (see Figure \ref{fig:relocated_hub}). Utilizing the population data, we identified our fourth hub in a similar southwestern region. However, this represented a sub-optimal placement, yielding a 10.8\% improvement to 6.785 km (see Figures \ref{fig:population_hubs} and \ref{fig:hub_comparison}). A summary of our results is presented in Table \ref{tab:summary_of_results}.

% \begin{figure}[htbp]
% \centerline{\includegraphics[width=\linewidth]{distribution comparison.png}}
% \caption{Comparison of delivery distances from the nearest hub before and after adding the fourth hub}
% \label{fig:distribution_comparison}
% \end{figure}

\section{Conclusion}

In this work, we presented an optimization method that can be generalized across various industries aiming to enhance the expansion or relocation of facilities. This method can be directly applied to any network comprising demand points and facilities. Many approaches in the literature rely on Euclidean distances in a two-dimensional space, which may underestimate actual distances. Furthermore, applications of the conditional \( p \)-median are relatively scarce in the literature. Our work seeks to inspire operational methods within the industry that could contribute to reducing carbon emissions in developing countries.

A promising area for future research could involve the application of the capacitated \( p \)-median problem, where each facility is assigned a capacity for the demand points it can serve. Additionally, a study aimed at quantifying the actual impact of optimizing facilities on indigenous carbon emissions presents another valuable direction for future investigation.

\section*{Acknowledgment}

We acknowledge the support provided by M\&P Logistics for their active collaboration in making this research possible through their datasets.

\bibliographystyle{IEEEtran} % or any other style you prefer
\bibliography{mybib} % Assuming your .bib file is named 'mybib.bib'

\end{document}